\documentclass[twocolumn]{aastex63}
\usepackage[T1]{fontenc}
\usepackage[all]{hypcap}
 
\usepackage{amsmath}

\graphicspath{{./}{figures/}}

\begin{document}

\title{
Abundance Diagnostics from Slitless Imaging Spectrometer: A Proof-of-Concept for MaGIXS-2
}

\newcommand{\uah}{Center for Space Plasma and Aeronomic Research, University of Alabama Huntsville, Huntsville, AL, 35812, USA}
\newcommand{\msfc}{NASA Marshall Space Flight Center, Huntsville, AL, 35812, USA}

\correspondingauthor{Biswajit Mondal}
\email{biswajit70mondal94@gmail.com, biswajit.mondal@nasa.gov}
\author[0000-0002-7020-2826]{Biswajit Mondal}
\affiliation{NASA Postdoctoral Program, NASA Marshall Space Flight Center, ST13, Huntsville, AL, USA}
\author[0000-0002-5608-531X]{Amy R. Winebarger}
\affiliation{\msfc}
\author[0000-0002-4454-147X]{P. S. Athiray}\affiliation{\uah}\affiliation{\msfc}

\begin{abstract}

Elemental abundance diagnostics in the solar corona are crucial for understanding energy transport, plasma heating, and magnetic activity. Most earlier imaging-spectroscopic studies have relied on slit-based spectrometers, which offer high spectral resolution but are limited in spatiotemporal coverage and temperature diagnostics. In contrast, slitless spectrographs like the Marshall Grazing Incidence X-ray Spectrometer (MaGIXS) produce overlapping spatial-spectral data (overlappograms) to enable wide-field imaging spectroscopy with broader temperature coverage. However, extracting elemental abundances from overlappogram data remains inherently challenging due to spatial and spectral confusion. In this work, we present a proof-of-concept technique for elemental abundance diagnostics from overlappograms, demonstrating its feasibility with simulated MaGIXS-2 observations.

\end{abstract}

\section{Introduction}

Measuring the Sun’s elemental composition is crucial to understand how energy and mass are transported from the lower atmosphere to the outer layers. Although the outer atmosphere derives its mass from deeper regions, the elemental composition of the corona differs significantly from that of the photosphere. In magnetically closed-loop regions, such as solar active regions (ARs), elements with low First Ionization Potential (FIP) are preferentially enhanced by factors of 3–4, whereas high-FIP elements generally maintain their photospheric abundances~\citep{Feldman_1992PhyS...46..202F, Baker_2013ApJ...778...69B, Zanna_2022ApJ...934..159D, Mondal_2023ApJ...955..146M}. This compositional anomaly — referred to as the FIP effect — is not spatially uniform and varies across different solar structures~\citep{Feldman_1993ApJ...414..381F,Widing_1997ApJ...480..400W,Brooks_2017NatCo...8..183B,Doschek_2019ApJ...884..158D,Vadawale_2021ApJ...912L..12V}. As a key diagnostic in heliophysics, the FIP effect offers critical insights into the mechanisms of coronal heating, plasma dynamics, and energy transport within the solar atmosphere.

Elemental abundances in the solar atmosphere are commonly derived from extreme ultraviolet (EUV) and X-ray spectroscopic observations. 
Several space experiments, including CubeSats and satellite missions, have determined elemental abundances using Sun-as-a-star observations, without requiring detailed spatial information~\citep[see, for instance,][]{sylwester_2013, sylwester_2014, Narendranath_2014, Brooks_2017NatCo...8..183B, Narendranath_2020, Dennis_2015, mondal_2021, Suarez_2023, Ng_2024}. Among these, the Solar X-ray Monitor (XSM;~\citealp{Mithun_2020SoPh..295..139M, Mithun_2022ApJ...939..112M, Vadawale_2021ApJ...912L..13V}) onboard Chandrayaan-2 has been extensively utilized to derive spatially integrated elemental abundances in different solar features by observing the Sun during the minimum phase of solar cycle 24, when the X-ray emissions are dominated by a single coronal feature. For example, \citet{mondal_2021, Mithun_2022ApJ...939..112M, Nama_2023SoPh..298...55N, rao_2023ApJ...958..190R} derived the time evolution of abundances during solar flares of different classes, while \citet{Zanna_2022ApJ...934..159D, Mondal_2023ApJ...955..146M} measured the temporal evolution of abundances during the lifetimes of active regions (AR:~\citealp{Mondal_2025ApJ...980...75M}). Similarly, \citet{Vadawale_2021ApJ...912L..12V} investigated the time evolution of abundances in all X-ray bright points (XBP:~\citealp{mondal_2023b}) present on the solar disk.

%




Slit-based EUV spectrometers, such as the EUV Imaging Spectrometer (EIS:~\citealp{Culhane_2007SoPh..243...19C}) onboard Hinode, have been widely used to produce spatially resolved abundance maps of ARs (e.g.,~\citealp{Baker_2013ApJ...778...69B,Zanna_2014A&A...565A..14D}), and solar flares~\citep{Andy_2024A&A...691A..95T}. These instruments acquire high-resolution spectral data along a narrow slit, providing spatial information in only one dimension at a time. To construct a two-dimensional field of view (FOV), the slit must be moved across the solar surface in a process known as rastering. This approach is inherently time-consuming, and the resulting 2D maps often mix spatial structures with temporal evolution. Moreover, most EUV spectrometers are primarily sensitive to cooler and warm plasma, limiting their ability to capture the hotter coronal components that are more effectively observed in the soft X-ray regime.

In contrast, slitless spectrographs such as the Marshall Grazing Incidence X-ray Spectrometer (MaGIXS;~\citealp{Champey_2022JAI}) overcome the limitations of slit-based instruments by providing simultaneous imaging and spectral information across an extended field of view, while also offer high temperature plasma diagnostics \citep{Athiray_2019ApJ,Mondal_2024ApJ...967...23M}. The data from these instruments, often referred to as ``spectroheliograms" or ``overlappograms", capture spectral data from the entire region of interest in a single exposure. The two recent flights of the MaGIXS sounding rocket experiment — MaGIXS-1 in 2019 \citep{Sabrina_2023ApJ} and MaGIXS-2 in 2024 — successfully demonstrated the capability to observe a variety of coronal structures using high-temperature diagnostic emission lines in the soft X-ray wavelength range.

Due to the nature of slitless observations, overlappogram data exhibit overlapping spectral and spatial information, making their interpretation challenging. Significant efforts have been made to develop robust unfolding methods to analyze and better interpret slitless spectroscopic data. For example, \citet{Winebarger_2019ApJ...882...12W} and \citet{Cheung_2019ApJ...882...13C} introduced a regularization-based inversion technique to decompose such data and retrieve key physical parameters. \citet{Winebarger_2019ApJ...882...12W} demonstrated the method’s effectiveness in deriving temperature maps and generating spectrally pure images of various ions under the assumption of fixed elemental abundances. More recently, \citet{Athiray_2025ApJ...980..100A} explored parameter optimization for the inversion of MaGIXS-1 data, also assuming known abundances. However, the inherent complexity of slitless data presents a challenge for simultaneously determining both temperature and elemental composition.

In this work, we present a proof-of-concept technique for deriving elemental abundances from overlappogram observations by simultaneously inverting for both temperature and abundances. We demonstrate the feasibility of this approach using simulated data of MaGIXS-2 instrument.

The rest of the paper is organized as follows: Section~\ref{sec-method} describes the detailed methodology, and Section~\ref{sec-results} presents the results. Section~\ref{sec_summary} provides a brief summary of the work.

\section{Method}\label{sec-method}

To invert the overlappogram data from MaGIXS-1, \citet{Sabrina_2023ApJ,Athiray_2025ApJ...980..100A} adopt the general inversion framework for spectral decomposition described by \citet{Cheung_2019ApJ...882...13C} and \citet{Winebarger_2019ApJ...882...12W}. They formulate the inversion problem as a set of linear equations:
\begin{equation}\label{eq1}
y = \sum_{\theta,k} M(F_{\theta}, T_k) \cdot EM({F_{\theta}, T_k})
\end{equation}
Here, \( y = (y_1, y_2, \ldots, y_{npix}) \) is the observed intensity in a single detector row along the dispersion direction, where $npix$ is the number of pixels on the detector.  \( EM({F_{\theta}, T_k}) \) is a linearized, one-dimensional array of emission measures for different field angles (\( F_\theta \)) and temperatures (\( T_k \)). The size of $EM$ is the total number of free parameters in the inversion, $nfree$.  \( M \) is the instrument response matrix ($npix \times nfree$), which maps the emission measure from each field angle and temperature onto detector pixels, producing a corresponding electron signal. In this formulation, the elemental abundances are assumed to be known and fixed within \( M \). 
The inversion of MaGIXS-1 flight data reported in \citet{Sabrina_2023ApJ} was found to best match with the coronal abundances by comparing multiple inversions of different fixed abundance models—namely coronal, hybrid, and photospheric. However, they did not allow the FIP bias to vary freely, which limited their ability to determine the spatial variation of the FIP bias.


In the present work, we relax this assumption and consider that \( M \) also depends on elemental abundances. Equation~\ref{eq1} then becomes:
\begin{equation}\label{eq2}
y = \sum_{\theta,k,Z} M(F_{\theta}, T_k, Z) \cdot EM(F_\theta,T_k,A_Z)
\end{equation}
Here, \( A_Z \) is the abundance of an element with atomic number \( Z \), whose emission lines fall within the spectral bandpass of the instrument.

\subsection{Abundance Diagnostic for MaGIXS}\label{sec-diagonostic}

\begin{figure}[ht!]
\centering
\includegraphics[width=1\linewidth]{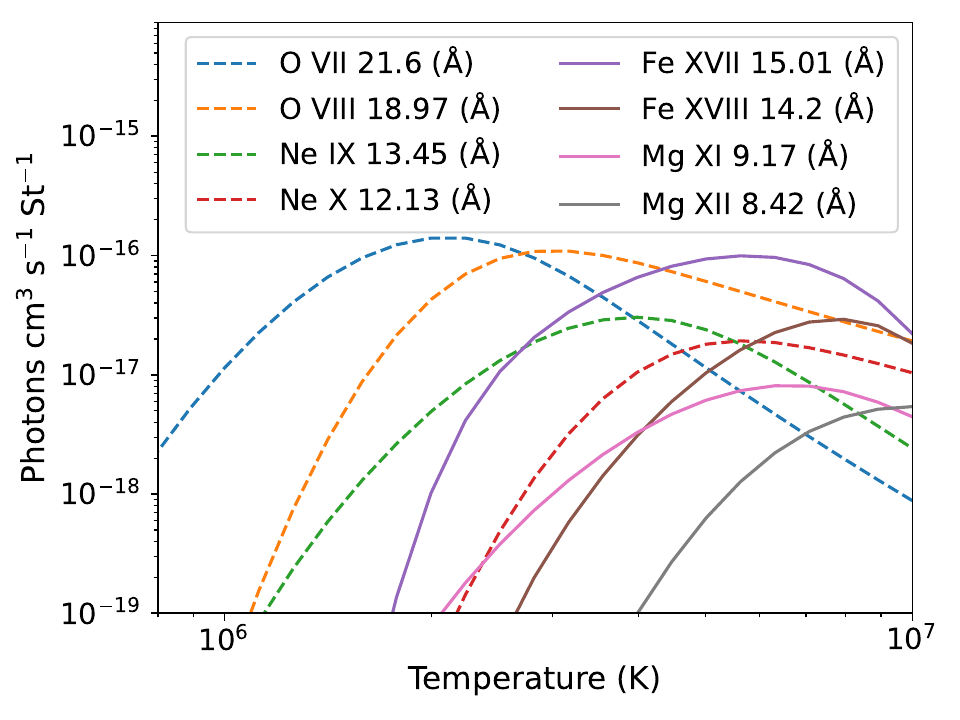}
\caption{Emissivity function of a few emission lines in MaGIXS passband. Dashed and solid lines represent high and low FIP elements respectively.}
\label{fig-emissivity}
\end{figure}

Considering Eq.~\ref{eq2}, the dimensionality of \( M \) becomes quite large, depending on the number of elements included in the abundance diagnostics. In the MaGIXS passband at typical AR temperatures, the high-FIP elements include O (VII and VIII) and Ne (IX and X), while the low-FIP elements include Fe (XVI - XIX) and Mg (XI and XII). Among these, the strongest spectral lines are from O and Fe. Figure~\ref{fig-emissivity} show the primary emissivity functions for a few lines in MaGIXS passband for photospheric abundances~\citep{Asplund_2021A&A...653A.141A}.

Therefore, instead of solving Eq.~\ref{eq2} for each elemental abundance separately, we group the elements by their FIP characteristics. Specifically, we treat the response of all low-FIP elements as one component and that of high-FIP elements as another. Solving the equation in this way yields two solutions of \( EM \) for each field angle and temperature: one corresponding to high-FIP elements and the other to low-FIP elements.
In this case, Eq.~\ref{eq2} becomes Eq.~\ref{eq3}.
\begin{equation}\label{eq3}
y = \sum_{\theta,k} {\Big(}M_h(F_{\theta}, T_k) \cdot EM_h(F_\theta,T_k) + M_l(F_{\theta}, T_k) \cdot EM_l(F_\theta,T_k){\Big)}
\end{equation}
where the indices \( h \) and \( l \) represent the high- and low-FIP components, respectively.

The ratio of the low-FIP solution to the high-FIP solution provides an estimate of the relative FIP bias. However, since the solution at each field angle is temperature-dependent, we do not take a direct ratio of the total solutions. Instead, we synthesize the emission of the strongest lines (e.g., Fe XVII) within the MaGIXS passband using both solutions, and then compute their ratio to estimate the FIP bias at each field angle. This approach allows us to ignore temperature ranges where MaGIXS has limited sensitivity and where the inversion result is subject to larger uncertainties.

In the following section, we demonstrate this method using simulated MaGIXS-2 data. We simulate the MaGIXS-2 observation using a DEM map derived from SDO/AIA observations, synthesize the MaGIXS-2 detector response using a known abundance map , perform the inversion using Eq.~\ref{eq2}, and then compare the recovered FIP bias with the known input value.

\subsection{Simulated MaGIXS-2 Observation}\label{sec_mag_synthetic}

{  To simulate the MaGIXS-2 data, we start with observations of the Sun from May 13, 2024 at 17:00 UT made by the Atmospheric Imaging Assembly (AIA; \citealt{lemen2012}).} We choose this day because there were several active regions across the solar disk, similar to the Sun on the day of the MaGIXS-2 launch.  
We downloaded AIA level-1 data from the Joint Science Operations Center (JSOC) and processed it to level 1.5 using standard procedures available in SunPy\citep{sunpy_community2020}. The data were then rebinned to match the MaGIXS plate scale of 2.84$^{\prime\prime}$ and convolved with a Gaussian point spread function (PSF) of $\sim$ 18$^{\prime\prime}$, consistent with the MaGIXS instrument resolution.

\begin{figure*}[ht!]
\centering
\includegraphics[width=1\linewidth]{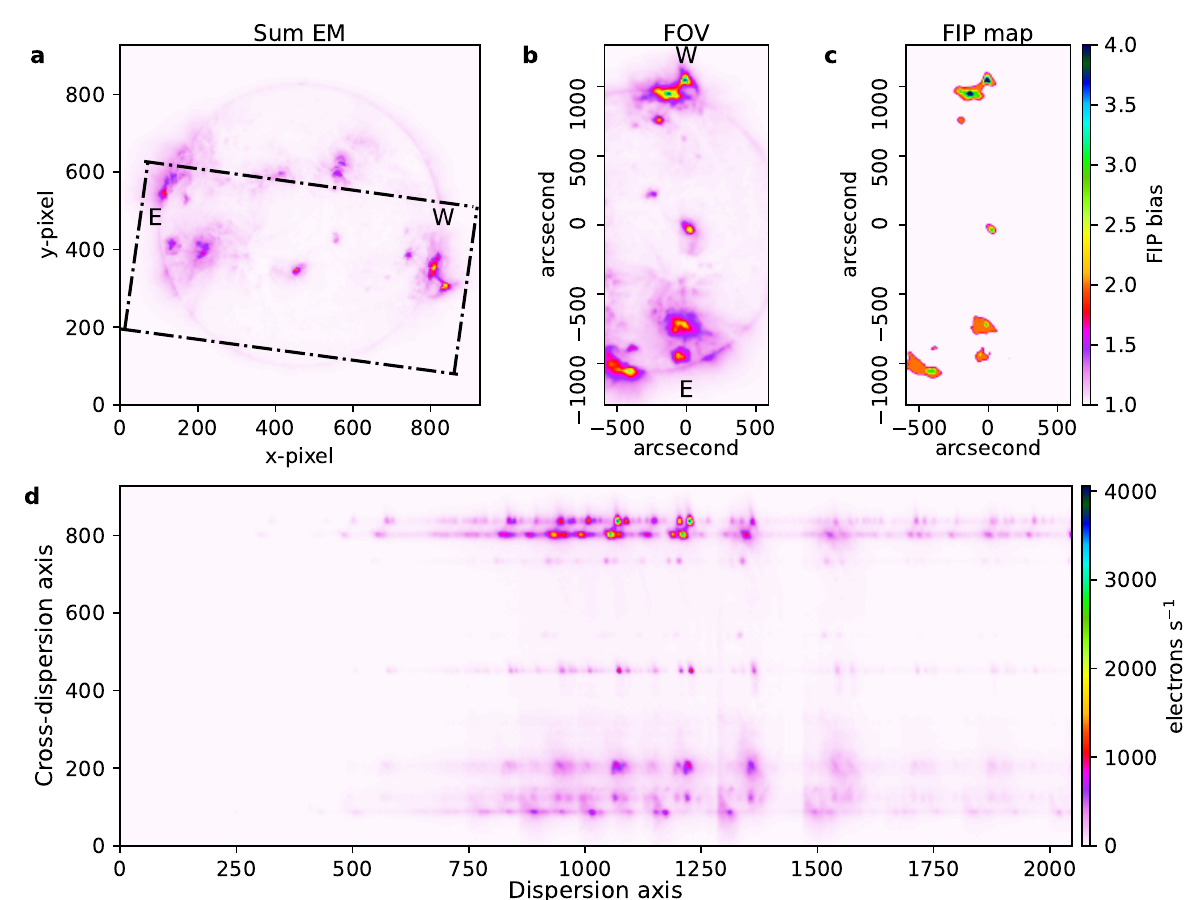}
\caption{Panel (a) shows the full-Sun emission measure (EM) map on May-13-2024 at 17:00 UT, derived from AIA observations and summed over temperature. The black box indicates the MaGIXS FOV. Panel (b) presents a vertically rotated view of the MaGIXS FOV from panel (a). Panel (c) shows the assumed FIP bias map, based on EM thresholds as described in the text. Panel (d) displays the simulated MaGIXS detector signal.}
\label{fig-Mag2Synthetic}
\end{figure*}

{  Using the standard AIA differential emission measure (DEM) inversion code \citep{Cheung_2015ApJ...807..143C}, we generated a DEM map of the full Sun for the temperature range logT=5.8 to 7.0.} The total emission measure (EM), obtained by summing over all temperatures, is shown in Figure~\ref{fig-Mag2Synthetic}a. The black box indicates the approximate MaGIXS-2 field of view (FOV), and Figure~\ref{fig-Mag2Synthetic}b shows a vertically aligned cutout of this region.

The AIA EUV passbands are dominated by Fe lines, so the AIA data is insensitive to abundances.  To simulate a spatially varying elemental composition, then, we generated a FIP bias map over the entire FOV based on the total EM only. Specifically, a FIP bias of 4 is assigned to regions where the total EM exceeds 60\% of the maximum value; a bias of 3 is assigned where total EM falls between 40–60\%, a bias of 2 where total EM is between 20–40\%, and a unit FIP bias is assigned elsewhere. The resulting FIP bias map is shown in Figure~\ref{fig-Mag2Synthetic}c.  
{  This assumed FIP bias map is motivated by previous observations of FIP bias maps for ARs (e.g., \citealp{Baker_2013ApJ...778...69B,Baker_2015ApJ...802..104B}), where the FIP bias is higher in the brightest regions of the Hinode/EIS Fe XII intensity map.
According to FIP bias theory, the physical motivation is that the most intense hot/X-ray emissions originate from higher portions of coronal loops, which are expected to exhibit higher FIP bias compared to regions of lower intensity, where emissions arise from lower heights. However, previous observations also show that even in areas with small brightenings, regions with greater plasma supply still exhibit higher FIP bias.}
Note that though this FIP bias map is reasonable (typical FIP biases are in the range of 1-4 and are typically larger in AR), we are simply assigning these values based on total EM and have no way of knowing what the actual FIP bias in these structures are.

Using this FIP map and the MaGIXS-2 instrument response function, we forward-modeled the expected MaGIXS-2 detector signal, shown in Figure~\ref{fig-Mag2Synthetic}d. In the forward model, we assume photospheric abundances~\citep{Asplund_2021A&A...653A.141A} for high-FIP elements, while the abundances of low-FIP elements are enhanced from their photospheric values according to the assigned FIP bias at each solar location.  Finally, to generate realistic data, we add Poisson noise to the signal.  

\subsection{Inversion}\label{sec_inversion}

We perform the inversion of each synthetic detector data row using Eq.\ref{eq3}, employing the ElasticNet regularization technique, a widely used algorithm implemented in the \texttt{scikit-learn} library\footnote{\url{https://scikit-learn.org/stable/modules/generated/sklearn.linear_model.ElasticNet.html}} in Python.
ElasticNet is a linear regression model that enforces both sparsity and smoothness by minimizing the following loss function:

\begin{equation}
\frac{1}{2n} \min_x \left( \frac{1}{\sigma_{\text{pix}}} \| y - Mx \|_2^2 + \alpha \rho \|x\|_1 + 0.5\alpha(1 - \rho) \|x\|_2^2 \right)
\end{equation}

Here, \( \frac{1}{2n} \) is a normalization factor over the number of data points, and \( \sigma_{\text{pix}} \) represents the pixel-wise noise, used to normalize the residual term \( \| y - Mx \|_2^2 \).  
The parameter \( \alpha \) controls the overall strength of regularization, while \( \rho \) sets the balance between sparsity (enforced via the \( \ell_1 \) norm) and smoothness (enforced via the \( \ell_2 \) norm).  
The terms \( \|x\|_1 \) and \( \|x\|_2^2 \) thus promote sparse and smooth solutions, respectively.  In this work, we use $\alpha = 1e-3$ and $\rho = 0.6$, which we find allows the inversion to adequately match the simulated data, though we have not done a complete study to determine the optimal inversion parameters \citep[e.g.][]{Athiray_2025ApJ...980..100A}.

\begin{figure*}[ht!]
\centering
\includegraphics[width=1\linewidth]{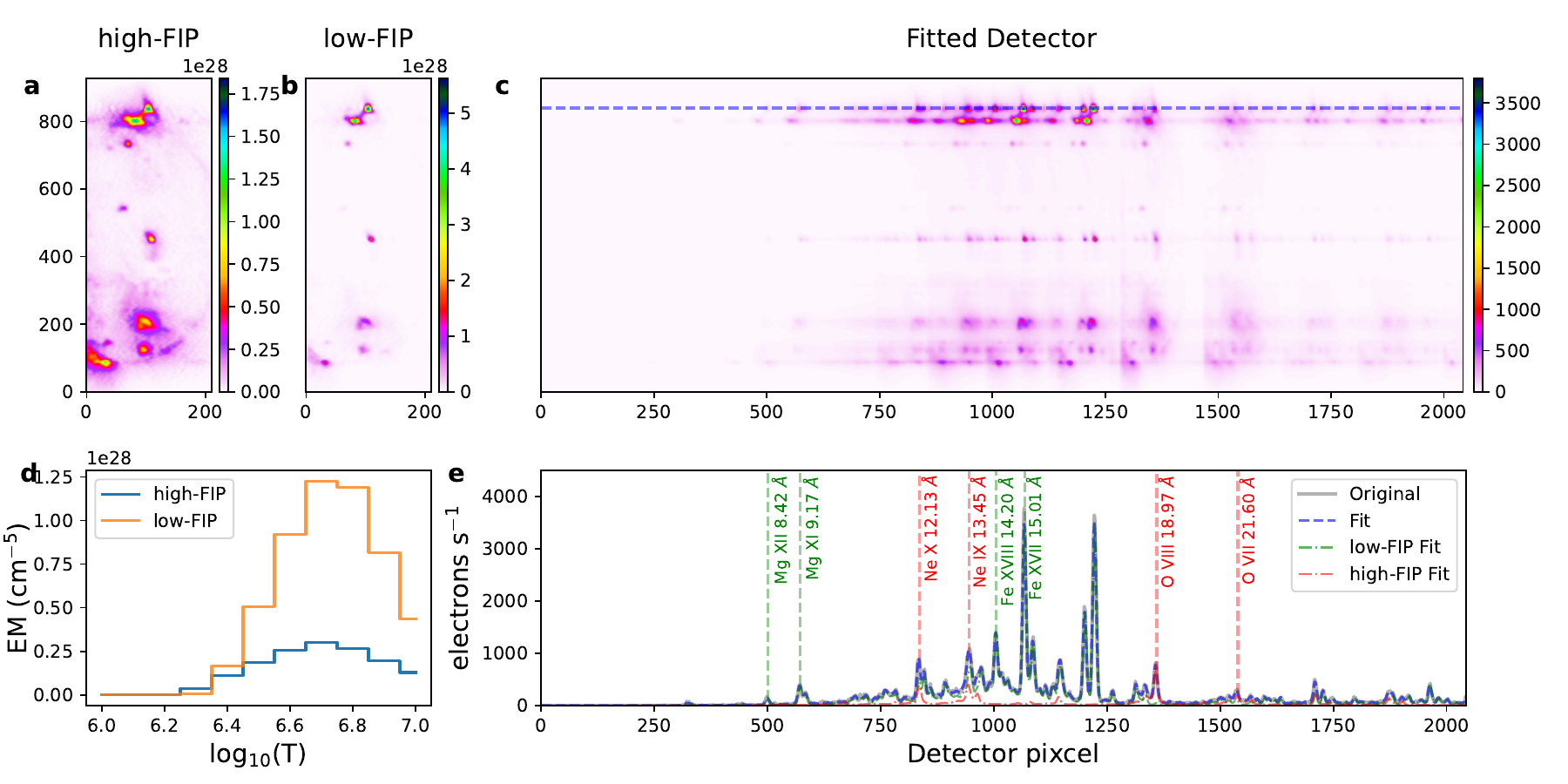}
\caption{Inverted solutions. Panels (a) and (b) show the total emission measure (EM) maps summed over temperature for the high-FIP and low-FIP solutions, respectively. Panel (c) shows the reconstructed detector image. Panel (d) presents a representative EM($T$) curve for a single pixel from panels (a) or (b) at (x, y) = (104, 840), illustrating the low- and high-FIP components. Panel (e) compares the recovered (blue) and true (gray) spectra along the horizontal blue dashed line in panel (c). The green and red curves represent the spectra corresponding to the low-FIP and high-FIP solutions, respectively.}
\label{fig-inv}
\end{figure*}

Figure~\ref{fig-inv} shows the inversion results. Panels (a) and (b) display the inverted emission measure summed over temperature for the low-FIP and high-FIP components, respectively. A representative EM curve for a single pixel is shown in panel (d), where the enhanced FIP bias results in a higher low-FIP solution compared to the high-FIP one. Panel (c) shows the inverted MaGIXS detector signal, which closely matches the actual detector signal shown in Figure~\ref{fig-Mag2Synthetic}d.
Panel (e) presents the spectrum along the horizontal dashed line marked in panel (c). The gray and blue curves represent the actual and fitted spectra, respectively, while the green and red curves correspond to the low-FIP and high-FIP components.
From the inverted low-FIP and high-FIP solutions, the elemental abundances are derived and discussed further in Section~\ref{sec-results}.

\section{Results}\label{sec-results}


The inversion discussed in Section~\ref{sec_inversion} yields two EM cubes (Figure~\ref{fig-inv}): one associated with low-FIP elements and the other with high-FIP elements.
The ratio between the low-FIP and high-FIP EM solutions provides a measure of the FIP bias. However, since the EM solutions are functions of temperature, we derive the intensity of the Fe XVII emission line—one of the strongest lines in the MaGIXS passband—to obtain a meaningful average.
This approach ensures that any emission predicted at temperatures to which MaGIXS is not sensitive is effectively excluded. Finally, the FIP bias is computed as the ratio of the derived Fe XVII intensities for the low-FIP and high-FIP components.

\begin{figure*}[ht!]
\centering
\includegraphics[width=1\linewidth]{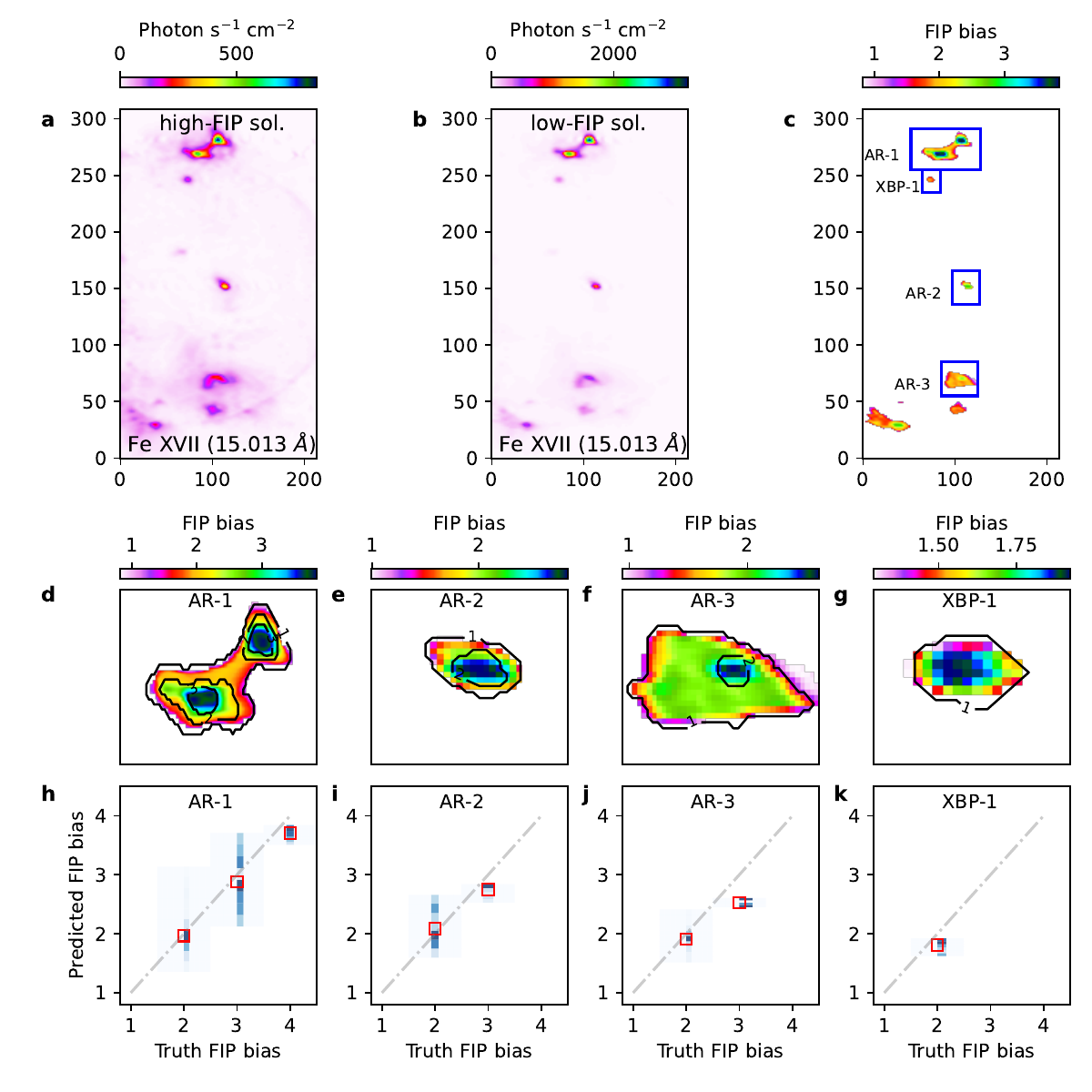}
\caption{Panels (a) and (b) show the Fe XVII intensity maps for the high- and low-FIP solutions, respectively, while panel (c) shows the ratio of these two solutions, representing the derived FIP bias map. Panels (d)–(g) present zoomed-in views of the regions marked by blue boxes in panel (c). The true FIP bias is overplotted with black contours. Panels (h)–(k) display 2D histograms corresponding to panels (d)–(g), comparing the predicted FIP bias with the true values. The red squares indicate the average predicted FIP bias, and the gray dashed lines represent the one-to-one correspondence with the true values.}
\label{fig-fipmap}
\end{figure*}

Figure~\ref{fig-fipmap}a and b show the Fe XVII intensity maps for the low-FIP and high-FIP solutions, respectively, while panel (c) presents their ratio. 
{  This ratio represents the derived FIP bias map, which we now compare with the true FIP bias map, shown in Figure~\ref{fig-Mag2Synthetic}c.} Since the FIP bias is calculated from the ratio of Fe XVII intensities, the measurement is only reliable in regions where the Fe XVII signal is strong. Therefore, the FIP bias map in panel (c) is shown only for regions with sufficient signal.
We have marked three active regions (ARs) and one X-ray bright point (XBP) with blue boxes in panel (c). These are the regions for which we will quantitatively compare the measured FIP bias with the true values. Panels (d)–(g) show zoomed-in views of these marked regions, with contours of the true FIP bias overplotted. The inversion successfully recovers the FIP bias structure in all these regions, closely matching the true values.
To quantify this agreement, panels (h)–(k) present 2D histograms comparing the true and predicted FIP bias values for all spatial locations in the corresponding sub-regions (d)–(g). The red squares indicate the average values, and the dashed lines represent the one-to-one correspondence with the true values.

We find that the average measured FIP biases closely match the true values. Some spread is observed in the measurements across different spatial locations, which arises from a combination of noise introduced by the inversion process and, in part, from the Poisson statistics of the simulated observations. As expected, the uncertainties are larger in regions where the signal is weaker. The uncertainty in the measured FIP bias is found to be within 18\%.
{  We also repeat the experiment for synthetic data without applying any FIP bias, and in this case, as expected, the inversion predicts an FIP bias close to unity everywhere within its uncertainty, which further validates this methodology.}

\section{Summary and Discussion}\label{sec_summary}

In this study, we presented and validated a novel methodology for deriving FIP-bias map from observations obtained by a slitless imaging spectrometer. We tested the method using synthetic MaGIXS data generated from AIA observations and a predefined spatially varying FIP-bias map. The inversion technique separates the contributions of low- and high-FIP elements by treating them as distinct emission components and applies ElasticNet regularization to solve for the emission measure (EM) distributions.
To quantify the FIP bias, we derived the Fe XVII line intensity from the inverted EM cubes, allowing us to compute a spatially resolved FIP bias map. Comparison with the input “truth” map shows that the inversion accurately reproduces both the large-scale and localized abundance structures, with an uncertainty within 18\%. The method's robustness was further demonstrated across several solar features, including active regions and X-ray bright points.

This technique provides a promising tool for future solar missions employing slitless spectrographs. In future work, we plan to apply this methodology to real MaGIXS-2 observations. 
Looking ahead, this methodology will be valuable for upcoming planned slitless spectrographs for solar studies, including NASA’s CubeSat Imaging X-ray Solar Spectrometer (CubIXSS;~\citealp{Caspi_2023SPD....5420704C}) and potential small explorer (currently in Phase A) the EUV CME and Coronal Connectivity Observatory (ECCCO;~\citealp{kathy2022}).

It should be noted that the performance of this method depends on the specifications of the instrument—such as spatial resolution, energy range, and the degree of spectral-spatial overlap—all of which influence the quality of the inversion. Therefore, the methodology should be adapted to the specific characteristics of each instrument.
For example, if a future instrument captures multiple low- and high-FIP elements with strong signal-to-noise ratio, or includes multiple emission lines from the same element across different ionization stages, then instead of grouping the instrument response by low- and high-FIP elements, it can be grouped by individual elements or by elements with emission lines formed in similar temperature ranges. In such cases, it would be possible to measure element-specific FIP biases.


\acknowledgments{
We acknowledge the helpful comments from an anonymous reviewer to improve the manuscript.
BM's research was supported by an appointment to the NASA Postdoctoral Program at the NASA Marshall Space Flight Center, administered by Oak Ridge Associated Universities under contract with NASA. We acknowledge the Marshall Grazing Incidence X-ray Spectrometer MaGIXS-2 team for making 
the instrument specific parameters accessible through NASA Low Cost Access to Space
(LCAS) program, funded via NASA Research Announcement (NRA) NNH21ZDA001N-HLCAS}

{\textit{Facilities:} MaGIXS-2, SDO(AIA).}

{\textit{Software:} Astropy~\citep{Astropy2018AJ....156..123A}, IPython~\citep{ipython_2007CSE.....9c..21P},matplotlib~\citep{matplotlib_2007CSE.....9...90H}, NumPy~\citep{2020NumPy-Array}, scipy~\citep{2020SciPy-NMeth}, SunPy~\citep{sunpy_community2020}, SolarSoftware~\citep{Freeland_1998}, 
Chianti~\citep{Dere_1997A&AS..chianti,chiantiV10_Zanna2020}}

\newpage
\bibliography{MaGIXS_Abund}   
\bibliographystyle{aasjournal}

\end{document}